\documentclass[english,]{lni} 

\usepackage{graphicx}
\usepackage{enumitem}
\usepackage{multicol}
\usepackage{multirow}
\usepackage{anyfontsize}
\usepackage{scalerel}

\usepackage{tikz}
\usetikzlibrary{svg.path}

\definecolor{orcidlogocol}{HTML}{A6CE39}
\tikzset{
    orcidlogo/.pic={
        \fill[orcidlogocol] svg{M256,128c0,70.7-57.3,128-128,128C57.3,256,0,198.7,0,128C0,57.3,57.3,0,128,0C198.7,0,256,57.3,256,128z};
        \fill[white] svg{M86.3,186.2H70.9V79.1h15.4v48.4V186.2z}
        svg{M108.9,79.1h41.6c39.6,0,57,28.3,57,53.6c0,27.5-21.5,53.6-56.8,53.6h-41.8V79.1z M124.3,172.4h24.5c34.9,0,42.9-26.5,42.9-39.7c0-21.5-13.7-39.7-43.7-39.7h-23.7V172.4z}
        svg{M88.7,56.8c0,5.5-4.5,10.1-10.1,10.1c-5.6,0-10.1-4.6-10.1-10.1c0-5.6,4.5-10.1,10.1-10.1C84.2,46.7,88.7,51.3,88.7,56.8z};
    }
}

\newcommand\orcidicon[1]{\href{https://orcid.org/#1}{\mbox{\scalerel*{
                \begin{tikzpicture}[yscale=-1,transform shape]
                \pic{orcidlogo};
                \end{tikzpicture}
            }{|}}}}

\begin{document}

\title[Large Language Models in Introductory Programming Education]{Large Language Models in Introductory Programming Education: ChatGPT's Performance and Implications for Assessments}


\author[Kiesler \& Schiffner]
{Natalie Kiesler \orcidicon{0000-0002-6843-2729}\footnote{DIPF Leibniz Institute for Research and Information in Education, Germany, \email{kiesler@dipf.de}}  \and
Daniel Schiffner \orcidicon{0000-0002-0794-0359}\footnote{DIPF Leibniz Institute for Research and Information in Education, Germany, \email{schiffner@dipf.de}}}
\startpage{1} 

\maketitle

\begin{abstract} 
This paper investigates the performance of the Large Language Models (LLMs) ChatGPT-3.5 and GPT-4 in solving introductory programming tasks. Based on the performance, implications for didactic scenarios and assessment formats utilizing LLMs are derived. For the analysis, 72 Python tasks for novice programmers were selected from the free site CodingBat. Full task descriptions were used as input to the LLMs, while the generated replies were evaluated using CodingBat's unit tests. In addition, the general availability of textual explanations and program code was analyzed. The results show high scores of 94.4 to 95.8\% correct responses and reliable availability of textual explanations and program code, which opens new ways to incorporate LLMs into programming education and assessment. 
\end{abstract}
\section{Introduction}
The advent of Large Language Models (LLMs), such as OpenAI's ChatGPT, Codex, and GitHub's Copilot, affects the educational landscape at its core, as LLMs offer entirely new possibilities, but also challenges for educators, learners, and institutions. Even though LLMs have only appeared very recently to a broader audience, research has started to address their implications on computing education, particularly programming. The generative potential may be used by educators for the design of new programming tasks \cite{Sarsa2022}, or for students to gather formative feedback \cite{kazemitabaar2023studying,zhang2022repairing}. At the same time, implications for programming pedagogy and assessments are being discussed \cite{Becker2023,bull2023generative,rudolph2023chatgpt}, as the low-threshold availability of LLMs raises new questions with regard to adequate task designs, students' contribution, plagiarism, and ethical conduct. Educators and institutions will soon need to reconsider the design of (formative) assessments. 
In this context, it is crucial to investigate the capabilities and limitations of LLMs for novice learners of programming, whose challenges have a well-documented history \cite{spohrer1986novice,McCracken2001amultinational,luxton-Reilly18}. 



The goals of this paper are (1) to investigate the potential of LLMs, such as ChatGPT, for the generation of correct, executable program code for introductory programming tasks, and (2) to discuss didactic scenarios including assessments for the use of LLMs in introductory programming education.

\section{Related Work on Large Language Models}
Research on Large Language Models (LLMs) started to increase ever since OpenAI's ChatGPT was launched with free access in November 2022. Other LLMs of interest for the context of computing and programming education comprise OpenAI's Codex, and GitHub's Copilot\footnote{available here: \url{https://openai.com/blog/openai-codex} and here: \url{https://github.com/features/copilot}}. 
Early papers on the performance of OpenAI's Codex conclude that the code generated by Codex outscores most students in CS1 \cite{finnie2022robots} and CS2 exercises \cite{Finnie2023}. 
Zhang et al. \cite{zhang2022repairing} found that Codex can help students to fix syntactic and semantic mistakes in their Python code. 
In a study with 69 programming novices, Kazemitabaar et al. \cite{kazemitabaar2023studying} explored the potential of Codex for solving programming tasks. Students with access to Codex had completed their tasks and increased their scores significantly, compared to those students without access to Codex.
GitHub's Copilot is also capable of successfully solving introductory programming tasks with few additional prompts or adjustments \cite{wermelinger2023usingcopilot,puryear2022githubcopilot}.

However, the code and feedback provided by LLMs along with usability issues imply that there is still room for improvements and further developments. 
For example, a study with 24 programmers concluded that the code generated by Copilot still contains errors \cite{vaithilingam2022expectation}. Even though test subjects preferred Copilot over the code completion plugin IntelliSense, longer code snippets were perceived as difficult to understand, edit and repair from the programmer's perspective. 
Similarly, Prather et al. \cite{prather2023s} identified challenges of novices when using Copilot, some of which are due to Copilot's design. 
A study by Denny et al. \cite{denny2023copilot} focused on the engineering of prompts 
when using Copilot to mitigate its performance deficits and identify prompts leading to better feedback and results. 


An exploratory interview study with five professional developers \cite{bull2023generative} summarizes opportunities and threats from an industry perspective, and discusses implications on software development education. Among them are scaffolding and fading of support based on Sweller's cognitive load theory~\cite{Sweller2011}, a change of assessment formats, and a
transitional period for novice learners before using LLMs. Rudolph et al. \cite{rudolph2023chatgpt} discuss ChatGPT's impact on traditional assessment formats in higher education. They recommend, e.g., fostering and assessing students' creative and critical thinking abilities, and letting students perform their competencies in class, or in authentic situations, while offering choices (if possible) to address students' interests. The conclusion is ``\textit{to help students learn how to use AI tools judiciously and understand their benefits and limitations}''~\cite{rudolph2023chatgpt}. This is also the focus of a recent working group within the context of the Innovation and Technology in Computer Science Education conference~\cite{prather2023wgabstract}.




\section{Methodology}
To address the current research desiderata on ChatGPT's performance in introductory programming tasks, and opportunities related to conversational programming in education and assessment, we define the following research question: \textit{How does a Large Language Model like ChatGPT perform when asked to solve introductory programming tasks?} 

To address this research question, all CodingBat tasks from Python code practice areas were used as input to GPT-3.5 (freely available), and GPT-4 (pay model) as the most recent versions of OpenAI's LLM. 
The CodingBat tasks are available within eight areas and contain simple, basic, and medium problems. Each area contains 6 to 12 exercises, resulting in a total of 72 tasks \cite{parlante2017codingbat}.
The advantage of CodingBat tasks is that they are available online, and have been authored by a computing educator from Stanford University. Moreover, the automatic evaluation of input via unit tests upon execution allows for a straightforward evaluation of the LLM's performance.
For every problem, the exact task description is presented to the LLM, (see \cite{denny2023copilot}).
Then the generated output is characterized using a simple metric (e.g., textual explanation, or program code contained).
If ChatGPT's response contained program code in Python, the code was used as input to CodingBat. Then the suggested program was executed, and the number of correct and incorrect test cases was recorded. 
In case ChatGPT could not provide a correct solution, additional prompts were explored as input and, once again, evaluated w.r.t. test case performance via CodingBat. 

\section{LLMs' Performance in Solving Introductory Programming Tasks}

\begin{figure}[t!]
    \centering
    \includegraphics[width=0.7\textwidth]{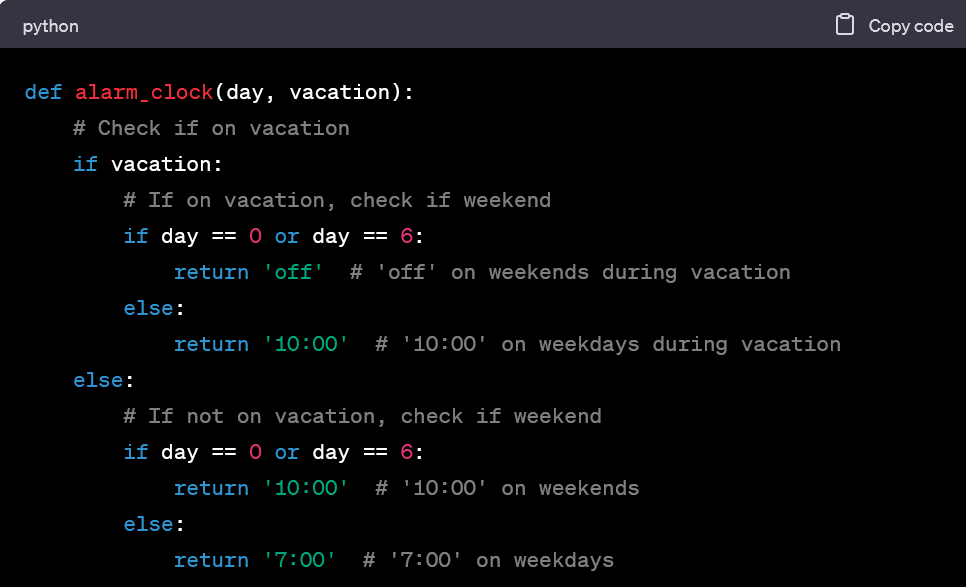}
    \caption{Code with comments for the $alarm\_clock$ task generated by ChatGPT-3.5}
    \label{fig:example_comment}
\end{figure}

We evaluated both Chat-GPT-3.5 and GPT-4 and their performance in introductory programming tasks provided by CodingBat~\cite{parlante2017codingbat}. Upon entering the full task description as an input, ChatGPT-3.5 immediately solved the task correctly in 69 out of 72 cases (95,8\%), while GPT-4 solved 68 out of 72 tasks correctly (94,4\%). Program code in Python was provided in all initial responses by both, ChatGPT-3.5 and GPT-4, even though the tasks do not explicitly ask for it. Textual explanations of the code were generated for 70 out of 72 tasks by both, ChatGPT-3.5 and GPT-4, and thus in 97,2\% of all test cases. Moreover, the code often contains explanatory comments (see Figure \ref{fig:example_comment}), or additional sample output is generated. 
Table \ref{tab:performance_results} presents the performance results of both LLM versions in each of the eight task areas. Thus, the availability of textual explanations, program code, and the generation of fully correct solutions passing all unit tests are summarized for the number of tasks in each area. 

\begin{table}[h!]
\fontsize{8pt}{8pt}\selectfont
    \centering
    \
    \caption{Summary of GPT-3.5's and GPT-4's performance in solving CodingBat tasks.}
    \label{tab:performance_results}
    \begin{tabular}{|p{1.4cm}|p{1.4cm}|p{1.2cm}|p{1.4cm}||p{1.5cm}|p{1.2cm}|p{1.4cm}|}
    \hline
        \textbf{CodingBat task area} & \textbf{GPT-3.5 textual explanation} & \textbf{GPT-3.5 program code} & \textbf{GPT-3.5 correct unit test results} & \textbf{GPT-4 textual explanation} & \textbf{GPT-4 program code} & \textbf{GPT-4 correct unit test results} \\ \hline
        \textbf{Warmup1} & 11/12 & 12/12 & 12/12 & 12/12 & 12/12 & 12/12 \\ \hline
        \textbf{Warmup2} & 9/9 & 9/9 & 9/9 & 9/9 & 9/9 & 9/9  \\ \hline
        \textbf{String1 } & 11/11 & 11/11 & 10/11 & 11/11 & 11/11 & 11/11 \\ \hline
        \textbf{List1 } & 12/12 & 12/12 & 12/12 & 11/12 & 12/12 & 12/12 \\ \hline
        \textbf{Logic1} & 8/9 & 9/9 & 8/9 & 9/9 & 9/9 & 9/9 \\ \hline
        \textbf{Logic2} & 7/7 & 7/7 & 6/7 & 6/7 & 7/7 & 6/7 \\ \hline
        \textbf{String2} & 6/6 & 6/6 & 6/6 & 6/6 & 6/6 & 4/6 \\ \hline
        \textbf{List2} & 6/6 & 6/6 & 6/6 & 6/6 & 6/6 & 5/6  \\ \hline
    \end{tabular}
\end{table}

The three errors made by Chat-GPT-3.5 were due to an additionally required method (task $make\_tags$), and ambiguity or a lack of clarity in the task description (task $near\_ten$, and $make\_bricks$). The four errors in the responses generated by GPT-4 were due to similar reasons: ambiguity in the task description (task $make\_chocolate$, and $count\_hi$). The reason why two other responses caused the feedback ``Bad code'' was that GPT-4 imported libraries, which is a general constraint among CodingBat tasks. These errors were observed for $count\_code$ and $centered\_average$.

\newpage
To improve the generated solutions, the following prompts were used: ``Please generate compilable Python code'' upon compile problems (task $make\_tags$), and ``The code fails the following test cases: <test cases>'' for failed test cases (e.g., for $near\_ten$). These prompts immediately resulted in an improved, correct response including an explanation. 
However, when asked to correct the response to the $make\_bricks$ task, ChatGPT-3.5 was reluctant to change its output. The prompt including the incorrect test cases had to be repeated three times, before resulting in a fully correct solution. This seemingly overconfidence was not observed with GPT-4. The latter immediately reached a fully correct solution after using the prompt with the failed test cases ($make\_chocolate$, and $count\_hi$), or the request to ``Please solve the task without an import.'' ($count\_code$ and $centered\_average$) once.

Despite high rates of successful task completion, the performance has to be discussed w.r.t. to several aspects. First of all, the selected tasks are straightforward and clear in most cases, as they were developed by an experienced educator. They contain little ambiguity and provide exemplary input and output.
It was thus a successful strategy to use the full task description as input to the LLM. Nonetheless, students should be aware that ambiguity in the task will likely cause incorrect responses, as ChatGPT may offer a solution to a
different problem. Now this does not mean that teachers should develop ambiguous tasks. On the contrary, clear assignments are still important so that students understand the task. 

Second, students need to adhere to general task constraints. For instance, many educators do not allow the use of libraries, or they provide certain function signatures for novice learners. The same is true for CodingBat tasks. Hence, students must be aware of such more or less explicit constraints, and provide them as additional information to the LLM, if they want to receive correct answers. Moreover, students need to know that ChatGPT can be overly confident, as it may not immediately change its proposed solution. We observed this phenomenon while using ChatGPT-3.5, but not with GPT-4. Providing failed test cases as a follow-up input seems to be an adequate strategy to improve the output, but fully trusting an LLM is not (yet) an option. 

A limitation of this work is due to the random nature of ChatGPT's responses. Thus, answers are arbitrary, and we doubt that the very same answers can be replicated by other researchers. Moreover, model solutions to all CodingBat tasks are available in GitHub repositories, so chances are ChatGPT was trained on such data. In addition, CodingBat only offers very small programs, with real novices as a target group, meaning that common second-semester tasks are very different, and so might be the LLM performance. Thus, continuous research on the evolving capabilities of LLMs is required to evaluate their implications on education for higher semesters.

\section{Discussion of Implications on Didactic Scenarios and \\ Assessments}

After having discussed ChatGPT's current performance (June 2023) in solving programming tasks, we now focus on the implications of these results on didactic scenarios and assessment formats in introductory programming courses. We, therefore, propose some exemplary settings for the inclusion of LLMs in learning activities and formative assessments.

As a guiding principle, we evaluated the results and didactic scenarios using the following rule: Assume that the response is invalid or contains errors and that the LLM may be overconfident and hallucinating. 
Considering the good performance for the analyzed tasks and having this rule in mind, it seems quite natural that current LLMs can be used by students for individual practice and self-assessment. 
The option to discuss a solution can, for example, provide valuable information to learners. This might be comparable to a peer-review session or direct input/feedback from an educator. Even though responses may not be perfect, LLMs can provide useful textual explanations and code suggestions students currently gather from other sources, e.g., Stack Overflow. 

As our examples showed high success rates for the given tasks, novice programmers could also use the generated code to compare it to their own solution. A simple request to compare solutions provides the rationale for why one solution may be preferable over another one. Even simple tasks can have multiple solutions, but it is not obvious to novice programmers when to choose one over the other. Interesting results from such an exercise may also be discussed further in class with additional elaboration by the facilitator.
 
Another concept is to let students generate multiple solutions by LLMs, such as ChatGPT on purpose, and to discuss them as part of a peer-group exercise w.r.t. advantages and disadvantages. GPT-3.5 and GPT-4 already generate different solutions in many cases, and the same is true upon regeneration of the reply. While they solve the given task, the discussion among peers gives students the option to understand the underlying problem class more thoroughly. In our experiments, GPT-4 created solutions that were more sophisticated and even more complex than necessary, for example, by including additional conditions (e.g., $same\_first\_last$). In other cases, GPT-3.5's responses were unnecessarily complex (e.g., $sum\_3$). Such discussions can potentially be very fruitful and help understand several problem-solving approaches for one problem. 

In another scenario, students' discussions with ChatGPT can be used for the evaluation of learning processes and students' mental models. Currently, we are limited to either a static/dynamic code analysis that builds upon a test-driven development approach. In contrast to that, strategies for problem-solving are rarely assessed. With an LLM like ChatGPT and the option to share such a discussion via a link, a more individualized approach is available, helping teachers to better understand issues novice programmers are facing. LLMs like ChatGPT thus enable a new form of reflecting on problem-solving approaches to programming tasks. The discussion with such a tool can provide insights into the logic used to solve a given task. It allows educators to see that and distinguish between algorithmic and coding issues, which are hard to detect with program code alone. 

Another formative assessment method may involve reflection exercises with a focus on critically discussing tasks and various program solutions with ChatGPT. This way, students can learn more about program code (i.e., develop knowledge about knowledge \cite{andersonKT2001}), while critically approaching LLMs and their generated solutions. 
Conversations with LLMs may therefore be used as an innovative assessment method for meta-cognitive competencies, that is the systematic approach towards solving (similar) problems~\cite{kiesler2020programmingcomp}. 
LLMs can further facilitate new ways to assess lower levels of cognitive complexity as it becomes easier to represent tasks aiming at the analysis and evaluation of code, which, in turn, can contribute to the development and transfer of problem-solving strategies. 
The same may apply to the understanding of seemingly simple programming concepts, or procedural knowledge (see \cite{kiesler_diss_2022,kiesler2020koli,kiesler2020zurarxiv} for the classification of programming competencies). 
A prerequisite for such an assessment is, of course, the identification of observable and reliable indicators for such a measurement, along with an introduction of learners to the potential and limitations of conversing with LLMs (see also \cite{bull2023generative,rudolph2023chatgpt}).

\section{Conclusion and Future Work}

In this study, we investigated the potential of large language models in introductory programming education and assessments. The research question addresses the performance of LLMs when asked to solve introductory programming tasks. To answer it, we utilized existing coding tasks, i.e., CodingBat, as input to ChatGPT-3.5 and GPT-4 to produce program code, which was then evaluated by CodingBat's unit tests. The results show a high success rate, ranging between 94.4 and 95.8\%, which allowed us to further discuss didactic scenarios including assessments. Several scenarios, which until recently have been limited to having an expert (educator or tutor) at a learner's side, seem more realistic now -- even for large courses. LLMs further enable the assessment of different cognitive process dimensions, such as understanding programming concepts, or reading and analyzing code, which are currently hard to implement, and rarely addressed in higher education programming courses. Overall, the availability of LLMs may be a blessing for education if used with caution and guidance. With this study, we showed that an application in the context of introductory programming tasks can be reasonable. We expect these tools to help novice programmers to better understand problems and concepts, and that we can overcome some of the current limitations in programming education and assessments. 

Options for future work are manifold, as the full potential and challenges of LLMs have not yet been evaluated. Several follow-up studies are currently work-in-progress. One of them is concerned with the evaluation of feedback types offered by LLMs~\cite{kiesler2023exploringthepotential}, and another one aims to investigate ChatGPT's performance in actual exam tasks of an introductory programming course. The development of benchmarks for certain tasks is also considered future work~\cite{prather2023wgabstract}, as this could help educators evaluate the adequacy of assessments and didactic scenarios.





\bibliography{lni-paper-example-de} 

\end{document}